\documentclass[twocolumn,prl,amsmath,amssymb,floatfix,superscriptaddress]{revtex4-1}
\usepackage{color}
\usepackage{graphicx}
\usepackage{dcolumn}
\usepackage{bm}
\usepackage{times}
\begin{document}


\def\qaf{${\bf Q}_{\rm AF}$}

\title{Temperature-dependent transformation of the magnetic excitation spectrum on approaching superconductivity in
Fe$_{1-x}$(Ni/Cu)$_x$Te$_{0.5}$Se$_{0.5}$}

\author{Zhijun~Xu}
\altaffiliation{ZJX and JSW contributed equaly to this work.}
\affiliation{Condensed Matter Physics and Materials Science
Department, Brookhaven National Laboratory, Upton, New York 11973,
USA} 

\author{Jinsheng~Wen}
\altaffiliation{ZJX and JSW contributed equaly to this work.}
\affiliation{Condensed Matter Physics and Materials Science
Department, Brookhaven National Laboratory, Upton, New York 11973,
USA}
\affiliation{Physics Department, University of California, Berkeley, CA 94720, USA}
\affiliation{Materials Science Division, Lawrence Berkeley National 
Laboratory, Berkeley, CA, 94720, USA}

\author{Yang~Zhao}
\affiliation{NIST Center for Neutron Research, National Institute of
Standards and Technology, Gaithersburg, Maryland 20899, USA}
\affiliation{Department of
Materials Science and Engineering, University of Maryland, College Park, 
Maryland, 20742, USA}
\author{Masaaki Matsuda}
\affiliation{Quantum Condensed Matter Division, Oak Ridge National Laboratory,
Oak Ridge, TN, 37831, USA}
\author{Wei Ku}
\author{Xuerong Liu}
\author{Genda~Gu}
\affiliation{Condensed Matter Physics and Materials Science
Department, Brookhaven National Laboratory, Upton, New York 11973,
USA}
\author{D.-H. Lee}
\author{R. J. Birgeneau}
\affiliation{Physics Department, University of California, Berkeley, CA 94720, USA}
\affiliation{Materials Science Division, Lawrence Berkeley National 
Laboratory, Berkeley, CA, 94720, USA}
\author{J.~M.~Tranquada}
\author{Guangyong~Xu}
\affiliation{Condensed Matter Physics and Materials Science
Department, Brookhaven National Laboratory, Upton, New York 11973,
USA}
\date{\today}

\begin{abstract}

Spin excitations are one of the top candidates for mediating electron
pairing in unconventional superconductors. Their coupling
to superconductivity is evident in a large number of systems,
by the observation of an abrupt redistribution of magnetic spectral weight 
at the superconducting transition temperature, $T_c$, for energies comparable 
to the superconducting gap. 
Here we report inelastic neutron scattering measurements 
on Fe-based superconductors, Fe$_{1-x}$(Ni/Cu)$_x$Te$_{0.5}$Se$_{0.5}$, that 
emphasize an additional signature.  The overall shape of the low energy magnetic
dispersion changes from two incommensurate vertical columns at $T\gg T_c$ to a 
distinctly different U-shaped dispersion at low temperature. Importantly, 
this spectral reconstruction is apparent for temperature up to $\sim 3T_c$.  
If the magnetic excitations are involved in the pairing mechanism, their 
surprising modification on the approach to $T_c$ demonstrates that strong 
interactions are involved.

\end{abstract}

\maketitle

In weak-coupling models of magnetically-mediated superconductivity, magnons 
essentially replace phonons as the pairing bosons~\cite{Scalapino2010}.  By 
assumption, the interaction between the electrons and bosons is not strong 
enough to modify the bosonic excitation spectrum.  In conventional systems, 
superconductivity does modify the self-energy of the phonons, but there is 
no significant change in the phonon dispersion~\cite{Weber2010}.  In many 
unconventional superconductors,  including high-$T_c$ 
cuprates~\cite{YBCO_resonance1,YBCO_Bourges,Dai_Resonance1,BSCCO_Keimer}, 
heavy Fermion superconductors~\cite{NKSato,Stock115}, and the recently 
discovered Fe-based superconductors~\cite{Christianson2008,
Chis2009prl,Qiu2009}, one observes, on cooling below $T_c$, the 
gapping of low-energy spin fluctuations and a shift of spectral weight to 
a ``resonance'' peak.
Empirically, the magnetic spectrum found above and below $T_c$ tends to be 
qualitatively the same.

\begin{figure}[b]
\includegraphics[width=\linewidth,trim=-15mm -17mm -3mm 15mm,clip]{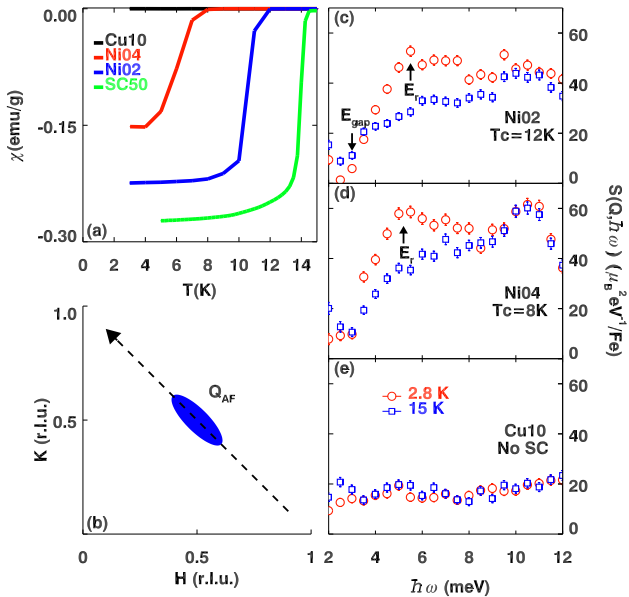}
\caption{(Color online) Magnetic susceptibility and inelastic neutron scattering measurements
performed on the Ni02, Ni04,  Cu10, and SC50 (FeTe$_{0.5}$Se$_{0.5}$)
samples. (a) Magnetic susceptibility measurements, and (c), (d), (e) magnetic 
neutron scattering intensity measured at \qaf\ with $T=2.8$~K (red circles) 
and 15~K (blue squares). The error bars represent the square root of the number
of counts. Fitted background obtained from constant-energy
scans has been subtracted from all data sets. The (HK0) scattering plane is 
plotted in (b) while the dashed line denotes the direction for the Q-scans 
shown in Fig.~\ref{fig:2}.} \label{fig:1}
\end{figure}

Here we study the low-energy spin fluctuations in single-crystal samples of the 
superconductor FeTe$_{0.5}$Se$_{0.5}$ (the ``1:1'' system, $T_c=14$~K) 
as we perturb the system 
by making partial substitutions for Fe.  Substituting 2\%\ and 4\%\ of Ni 
reduces $T_c$ to 12~K and 8~K, respectively, while 10\%\ of Cu results in 
an absence of superconductivity, as shown in Fig.~\ref{fig:1}~(a).  
Our inelastic neutron scattering measurements show that low energy ($\hbar\omega
\alt 12$~meV) magnetic excitations transform from having two peaks clearly away 
from the antiferromagnetic (AF) wave-vector at high temperature in the normal
state, to having a broad maximum near the   AF wave-vector at low temperature
in the superconducting phase. This drastic change on the magnetic dispersion
between the superconducting and non-superconducting phases suggests that strong
correlations between electrons have to be taken into account when the magnetic 
and electronic properties of the ``1:1'' system are considered. 

Single crystals of  Fe$_{1-x}$(Ni/Cu)$_x$Te$_{0.5}$Se$_{0.5}$ were grown 
by a unidirectional solidification method~\cite{Wen2011} at Brookhaven National 
Laboratory. The lattice constants are $a=b=3.81$~\AA, and $c=6.02$~\AA, using 
the two-Fe unit cell.  For convenience, we label these samples as Ni02, 
Ni04, and Cu10, according to the amount of Ni/Cu doping on the Fe site.
The neutron scattering experiments on the two Ni02 and 
Ni04 samples were carried out on the BT7 triple-axis-spectrometer at the NIST 
Center for Neutron Research, using beam collimations of 
open-$50'$-S-$50'$-$240'$, a fixed final energy of 14.7 meV and two 
pyrolytic graphite filters after the sample. The Cu10 sample was measured on 
the HB1 triple-axis-spectrometer at the High Flux Isotope Reactor, Oak Ridge 
National Laboratory. with beam collimations of $48'$-$40'$-S-$60'$-$240'$, 
fixed final 
energy of 13.5 meV , and two pyrolytic graphite filters after the sample. 
No static order around (0.5,0,0.5) was found in any
of the three samples. The inelastic scattering experiments were all 
performed in the $(HK0)$ zone, so that the scattering plane is defined by the
[100] and [010] wave-vectors. All data have been normalized
into absolute units of $\mu_B^2eV^{-1}$/Fe by incoherent
elastic scattering intensities from the samples. X-ray diffraction measurements 
of lattice parameters were performed at beamline X22B of the National 
Synchrotron Light Source, Brookhaven National Laboratory.

We are interested in the magnetic excitations near 
the AF wave-vector ${\bf Q}_{\rm AF}=(0.5,0.5,0)$.  
Figure~\ref{fig:1}~(c)-(e) shows the measured inelastic neutron scattering 
intensity as a function of energy obtained at $T=2.8$~K and 15~K for all three 
samples. It has been established in previous studies~\cite{Argyriou2009,Lee2010,
Lis2010,Zhijun_2011} that the unperturbed superconductor has a magnetic 
resonance peak at $E_r\sim 7$~meV.  Here we see that $E_r$ decreases to 
$\sim6$ and 5~meV in the Ni02 and N04 samples, respectively, while there is no 
observable resonance in the nonsuperconducting Cu10.  One can also see a spin 
gap of about 3~meV in Ni02, but the gap is more difficult to resolve for Ni04.

\begin{figure}[t]
\begin{center}
\includegraphics[width=\linewidth,trim=-5mm -5mm 0mm 5mm,clip]{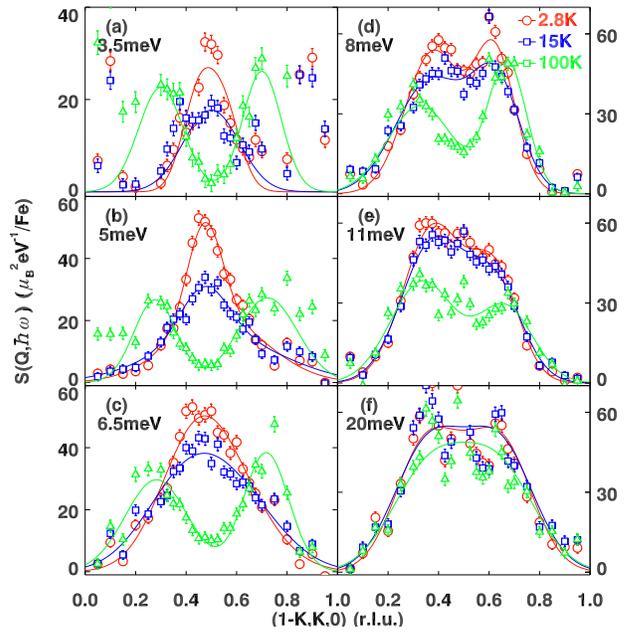}
\end{center}
\caption{(Color online) Wave-vector dependence of the magnetic scattering intensity along the 
transverse direction through \qaf\ [see Fig.~\ref{fig:1} (b)] 
for the Ni04 sample
at $T=2.8$~K (red circles), 15~K (blue squares), and 100~K (green triangles), 
obtained at excitation energies (a) 3.5~meV, (b) 5~meV, (c) 6.5~meV, (d) 8~meV, 
(e) 11~meV, (f) 20~meV [which was measured in a higher zone, near 
${\bf Q}=(1.5,0.5,0)$].  Solid lines are guides to the eye.} 
\label{fig:2}
\end{figure}

Things get more interesting when we look at the wave-vector ({\bf q}) dependence of the magnetic scattering.  It has been established in previous studies~\cite{Argyriou2009,Lee2010,Lumsden2010nf} of superconducting FeTe$_{1-x}$Se$_x$ that the magnetic excitations disperse from \qaf\ only in the transverse direction, along $[1,-1,0]$.  Figure~\ref{fig:2} shows scans along this direction for the Ni04 sample at a series of energies, illustrating the variation of the {\bf q} dependence as the temperature changes from 2.8~K ($\ll T_c$) to 15~K ($\gtrsim T_c$) and then up to 100~K ($T\gg T_c$).  The variations are minor at the higher energies, as in Fig.~\ref{fig:2}(e)-(f), but become dramatic for $E\sim E_r\approx 5$~meV, as in Fig.~\ref{fig:2}(a)-(c).  The change from $T\ll T_c$ to $T\gtrsim T_c$ is simply the standard resonance behavior.  The feature that we wish to emphasize is the change from a single commensurate peak at $T\gtrsim T_c$ to a pair of well-resolved incommensurate peaks at $T\gg T_c$.  This change cannot be confused with a temperature-dependent change in peak width.  

\begin{figure}[t]
\begin{center}
\includegraphics[angle=90,width=1.05\linewidth,trim=4mm 25mm 0mm 5mm,clip]{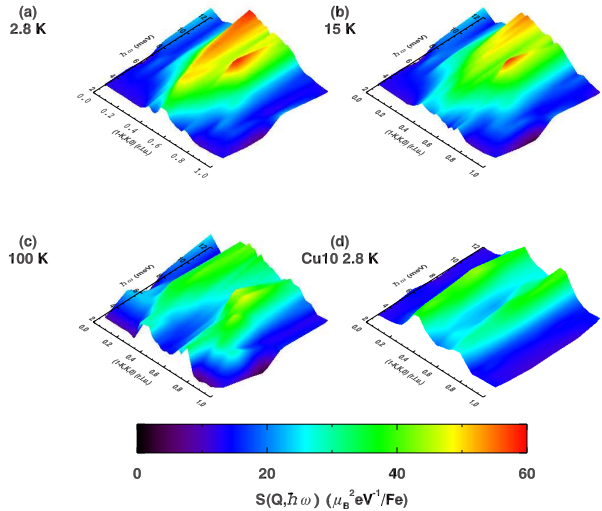}
\end{center}
\caption{(Color online) Magnetic scattering intensity plotted for the Ni04 sample
in energy-momentum space at (a) 2.8~K, (b) 15~K, and (c) 100~K. Results 
for the Cu10 sample measured at 2.8~K are plotted in (d). The data have been smoothed, and 
non-magnetic sharp spurious signals [see Fig.~\ref{fig:2}(a)] have been removed for 
better visual effects.} \label{fig:3}
\end{figure}

The same data are presented again, slightly cleaned up and in a different 
format, in Fig.~\ref{fig:3}(a)-(c).  The lower-temperature data exhibit a 
U-shaped dispersion, with the bottom of the U at $\sim E_r$.  Except for the 
change in the resonant peak, the basic shape of the dispersion does not really 
change on crossing $T_c$.  In contrast, the dispersion at 100~K is 
qualitatively different: it looks like the legs of a pair of trousers.  
It also looks very similar to the low-temperature dispersion of the 
non-superconducting Cu10 sample shown in Fig.~\ref{fig:3} (d).  

\begin{figure}[t]
\begin{center}
\includegraphics[width=\linewidth,trim=-20mm 15mm -5mm 18mm,clip]{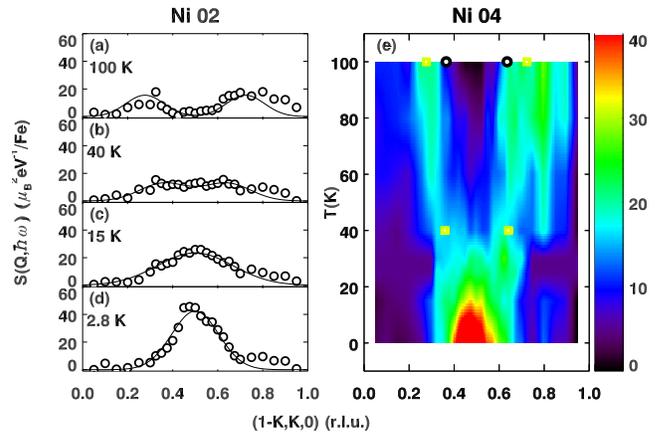}
\end{center}
\caption{(Color online) Thermal evolution of the magnetic scattering at $\hbar\omega=5$~meV. 
The data are measured through \qaf\ along the transverse direction for the Ni02 sample at 
(a) 100~K, (b) 40~K, (c) 15~K, (d) 2.8~K, and (e) for the Ni04 sample plotted as an intensity contour map in temperature--wave-vector space.
The data have been smoothed. 
The yellow and black symbols in (e) denote
the corresponding peak positions for the Ni02 
sample (yellow squares) and for a superconducting Fe$_{1+\delta}$Te$_{0.35}$Se$_{0.65}$ sample~\cite{Zhijun_2011}. 
} \label{fig:4}
\end{figure}

There is clearly a major change in the low-energy portion of the dispersion 
between 15 and 100~K, but how does it change between those temperatures?  
This is illustrated in Fig.~\ref{fig:4}.  Focusing in particular on the 
results for the Ni04 sample, in Fig.~\ref{fig:4}(e) we see that the crossover 
is continuous in temperature, but with a reasonably defined mid-point at 
$30\pm10$~K.  For Ni02, the midpoint may be closer to 40~K.  In both cases, 
the crossover occurs at temperatures of order $3T_c$.  We previously 
observed~\cite{Zhijun_2011} hints of this temperature dependent modification 
of the dispersion in superconducting FeTe$_{0.35}$Se$_{0.65}$; however, 
the high-temperature incommensurability was not as large nor as well resolved 
as for the Ni- and Cu-doped samples [see Fig.~\ref{fig:4} (e)].

It is possible to see the incommensurate columns of magnetic scattering even 
at low temperature when the superconductivity is suppressed, as shown for 
the Cu10 sample in Fig.~\ref{fig:3}(d).  A similar low-temperature spectrum has 
been observed previously in non-bulk-superconducting ``1:1'' samples such as 
Fe$_{1.04}$Te$_{0.73}$Se$_{0.27}$~\cite{Lumsden2010nf} and 
Fe$_{1.10}$Te$_{0.75}$Se$_{0.25}$~\cite{Swiss11}.  

\begin{figure}[t]
\begin{center}
\includegraphics[width=\linewidth,trim=-10mm 0mm -5mm 40mm,clip]{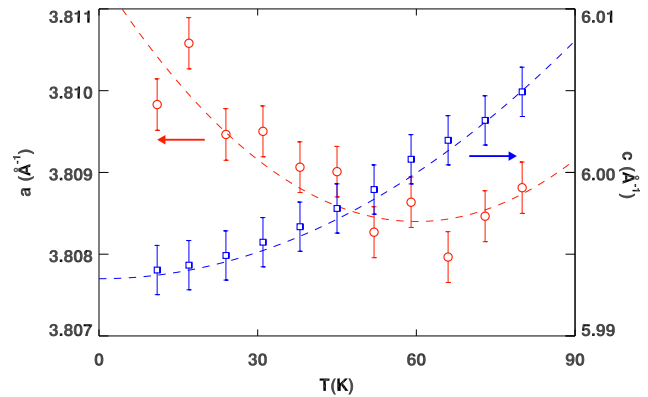}
\end{center}
\caption{(Color online) Lattice parameters $a$ (red circle) and $c$ (blue squares) 
measured on the Ni04 sample.} \label{fig:5}
\end{figure}

There is an evident pattern that superconducting 1:1 samples have commensurate 
or almost commensurate magnetic excitations at the resonance energy, while 
non-superconducting samples have incommensurate excitations.  Our results for 
the Ni-doped samples show that it is possible for a sample to transform from 
the incommensurate phase at high temperature to the low-energy-commensurate 
phase on cooling.  The commensurability appears at the energy scale of the 
resonance energy at a temperature of $\sim 3T_c$.  One possible explanation
of this incommensurate-to-commensurate transformation may be related to 
the strong orbital correlations in this system. In the iron-based 
superconductors, it has been proposed that there are competing electronic 
instabilities similar to those in the cuprates~\cite{DHLee,Zhai}.
In addition to  antiferromagnetism and superconductivity, the material also has 
a propensity toward $xz/yz$ orbital ordering. Such ordering can modify the 
shape of the Fermi surface (e.g.,  enlarging one electron Fermi pocket while 
shrinking the other), leading to enhancement of  the commensurate magnetic 
scattering between the hole and the electron pockets~\cite{Zhai}. Even if 
long-range orbital ordering does not set in upon cooling, the system may still 
have regions of locally ordered or slowly fluctuating orbital correlations. It 
is possible that the crossover we observe at $\sim3 T_c$ reflects such an 
orbital ordering ``transition'' in the presence of disorder. 
An additional experimental support of this scenario is given by 
x-ray scattering measurements on Fe$_{1+y}$Se$_{0.57}$Te$_{0.43}$~\cite{Gresty} 
where it was shown that a weak lattice distortion, reflecting  the breaking 
of 90 degree rotation symmetry,  sets in around 40~K,  similar to the crossover 
temperature observed in our experiment. In our samples, we did not see 
such a lattice distortion. Nevertheless, the temperature dependence of the
lattice parameters 
(Fig.~\ref{fig:5}) obtained from the Ni04 sample indicate an anomalous in-plane
expansion below 60~K, while  $c$ decreases monotonically
with cooling.

This temperature-dependent transformation of the magnetic spectrum is unusual 
among unconventional superconductors.  For example, in superconducting 
YBa$_2$Cu$_3$O$_{6+x}$ systems~\cite{YBCO_resonance1,YBCO_Bourges,
Dai_Resonance1}, the spin resonance develops at 
commensurate wave-vectors below T$_c$, but the ``hour-glass'' shaped dispersion
with a commensurate saddle point is still compatible with the high temperature
spectrum, within the increased q widths. In superconducting
La$_{2-x}$Sr$_x$CuO$_4$ the spin resonance occurs at lower energies where
the spin fluctuations are incommensurate~\cite{Christensen2004,Tranquada2004}, 
both in the normal and superconducting phases.   Returning to the analogy with 
electron-phonon coupling, strong interactions can lead to a modification of 
the spectrum through a structural phase transition, as 
occurs~\cite{Shirane1971} in Nb$_3$Sn at a temperature above the 
superconducting $T_c$.  In the present case, strong interactions appear 
necessary to cause the transformation from incommensurate to commensurate 
magnetic excitations. It is therefore reasonable to expect that measurements 
on the electronic structures in the system may also provide hints of this 
transformation in the same temperature range.  We note that a related 
precursor evolution of the charge response is apparent in the 
optical conductivity of 1:1 superconductor~\cite{HomesFeTe}.  It may also be 
worth noting that a thermally-induced enhancement of magnetic moments has been 
identified in non-superconducting Fe$_{1.1}$Te~\cite{Zaliznyak2011} at 
a similar temperature scale. 

Commensurate excitations are needed for the spin-fluctuation mechanism of 
electron pairing considered by a number of 
authors~\cite{Maier2009, Arita,Chubukov, TesanovicEPL, DHLee}.  In such a 
scenario, the momentum of the repulsive spin excitations couples the 
nearly-nested hole and electron pockets, and in turn allows a superconducting 
gap to develop on both sets of pockets, though with opposite phases.  
Obviously, losing the commensuration of the spin excitations would seriously 
impair the development of superconductivity in this kind of weak coupling 
scenario.  On cooling, do the electronic and magnetic correlations adjust 
themselves to enable the spin-fluctuation mechanism?  If so, what are the 
energetic tradeoffs associated with this transformation?  And can interactions 
strong enough to achieve this transformation lead to effectively the same 
pairing mechanism as the one identified from a weak-coupling approach?   
We hope that these questions will be addressed by future investigations.

\acknowledgements{We thank Igor Zaliznyak for useful discussions. Work at 
BNL is
supported  by the Office of Basic Energy Sciences, U.S. Department of Energy 
under contract No. DE-AC02-98CH10886.  Work at Berkeley is supported by the 
same office through contract No. DE-AC02-05CH11231.
The research at ORNL was sponsored by the Scientific User Facilities Division, 
Office of Basic Energy Sciences, U. S. DOE. }


%

\end{document}